\documentclass[useAMS,usenatbib]{mn2e}
\usepackage{aas_macros}
\usepackage[a4paper,centering, totalwidth=520pt, totalheight=700pt]{geometry}

\usepackage[fleqn]{amsmath} 
\usepackage{graphicx}
\usepackage{amssymb}
\usepackage{amsmath}
\usepackage{enumerate}
\usepackage{microtype}
\usepackage{color}
\usepackage{tabu}
\usepackage{multirow}
\usepackage{dcolumn}

\newcommand{\beq}{\begin{equation}}
\newcommand{\eeq}{\end{equation}}
\newcommand{\bal}{\begin{aligned}}
\newcommand{\eal}{\end{aligned}}
\newcommand{\beqa}{\begin{equation}\begin{aligned}}
\newcommand{\eeqa}{\end{aligned}\end{equation}}
\newcommand{\avg}[1]{\left\langle{#1}\right\rangle}

\newcommand{\hiMsun}{h^{-1} \rm M_\odot}

\newcommand{\Msun}{\rm M_\odot}

\newcommand{\Rfivec}{R_{\rm 500c}}
\newcommand{\Mgas}{M_{\rm g}}
\newcommand{\Mstar}{M_{\rm *}}
\newcommand{\Mbary}{M_{\rm b}}
\newcommand{\Mtot}{M_{\rm \Delta}}

\newcommand{\fgas}{f_{\rm g}}
\newcommand{\fstar}{f_{\rm *}}
\newcommand{\fbary}{f_{\rm b}}
\newcommand{\fb}{f_{\rm b}}
\newcommand{\fhot}{f_{\rm h}}
\newcommand{\fcold}{f_{\rm c}}
\newcommand{\fcs}{f_{\rm c*}}

\begin{document}

\title[Hot gas and galaxies in Rhapsody-G clusters]{Rhapsody-G simulations: galaxy clusters as baryonic closed boxes and the covariance between hot gas and galaxies}
\author[H.-Y. Wu et al.]{Hao-Yi Wu,$^{1}$
\thanks{Present address: California Institute of Technology, MC 367-17, 
Pasadena, CA 91125, USA. E-mail: hywu@caltech.edu}
August E. Evrard,$^{1}$
Oliver Hahn,$^{2}$
Davide Martizzi,$^{3}$ \newauthor
Romain Teyssier,$^{4}$
Risa H. Wechsler$^{5,6}$ 
\\
$^{1}$ Department of Physics, University of Michigan, Ann Arbor, MI 48109, USA\\  
$^{2}$ Department of Physics, ETH Zurich, CH-8093 Z\"urich, Switzerland \\
$^{3}$ Department of Astronomy, University of California, Berkeley, CA 94720-3411, USA\\
$^{4}$ Institute for Computational Science, University of Zurich, CH-8057 Z\"urich, Switzerland\\
$^{5}$ KIPAC, Physics Department, Stanford University, Stanford, CA 94305, USA\\
$^{6}$ SLAC National Accelerator Laboratory, Menlo Park, CA 94025, USA\\ 
}
\maketitle

\begin{abstract}
  Within a sufficiently large cosmic volume, conservation of baryons
implies a simple `closed box' view in which the sum of the baryonic
components must equal a constant fraction of the total enclosed mass.
We present evidence from {\sc Rhapsody-G} hydrodynamic simulations of
massive galaxy clusters that the closed-box expectation may hold to a
surprising degree within the interior, non-linear regions of haloes.  
At a fixed halo mass, we find a significant anti-correlation between hot
gas mass fraction and galaxy mass fraction (cold gas + stars), with a
rank correlation coefficient of $-0.69$ within $\Rfivec$.  
Because of this anti-correlation, the total baryon mass serves as a low-scatter
proxy for total cluster mass.  The fractional scatter of total baryon
fraction scales approximately as $0.02 (\Delta_c/100)^{0.6}$, while
the scatter of either gas mass or stellar mass is larger in magnitude
and declines more slowly with increasing radius.  We discuss potential
observational tests using cluster samples selected by optical and hot
gas properties; the simulations suggest that joint selection on
stellar and hot gas has potential to achieve $5\%$ scatter in total
halo mass.
\end{abstract}


\section{Introduction}

The abundance of galaxy clusters as a function of cluster mass is sensitive to both the growth of structure and cosmic expansion, providing not only stringent constraints on cosmological parameters but also consistency checks for the theory of gravity (see e.g. \citealt{Miller01,Vikhlinin09,Mantz10,Rozo10, Rapetti13,Benson13}; and \citealt{Allen11} for a review).  In cluster cosmology, the key is to accurately infer the mass of galaxy clusters from their observable properties, including gas mass and temperature from X-ray emission \citep[e.g.][]{Mantz14}, galaxy content from imaging and galaxy dynamics from spectroscopy \citep[e.g.][]{Kravtsov14,Mamon13}, and strong and weak gravitational lensing effects \citep[e.g.][]{vdLinden14}.  Each of these mass proxies exhibits a certain amount of scatter around the true mass; minimizing and characterizing this scatter is essential for precision cosmology from galaxy cluster surveys \citep[e.g.][]{LimaHu05,Wu09}.

To achieve accurate mass measurements, multi-wavelength observations have often been conducted for the same sample of clusters; for example, the CLASH project includes comparison between mass proxies from weak lensing, X-ray, and velocity dispersion \citep{Postman12,Donahue14,Biviano13}; clusters observed by the South Pole Telescope using the Sunyaev-Zeldovich effect have been followed up photometrically and spectroscopically \citep{Song12,Ruel14}.  When multiple mass tracers are available for the same sample of galaxy clusters, a joint selection can reduce the mass scatter.  In particular, the reduction of mass scatter is most effective when two mass tracers are anti-correlated with each other at a given mass \citep[e.g.][]{Cunha08,Stanek10, Rozo14b, Evrard14}.

Hydrodynamical simulations of galaxy clusters have been a powerful tool for understanding mass proxies  \citep[e.g.][]{Evrard96,Kravtsov06,Rasia06,Nagai07,Stanek10,Fabjan11,Rasia12,Angulo12,Saro12} and the evolution of gas and stellar mass in clusters \citep[e.g.][]{Kravtsov05,Ettori06,Puchwein10,Young11,Battaglia13,Planelles13}.  Recent results have shown that it is necessary to include the feedback of active galactic nuclei (AGN) in order to prevent catastrophic over-cooling in the cluster core, thus bringing the star formation in massive galaxies down to realistic levels and producing overall stellar mass fractions in better agreement with observations \citep[e.g.][]{Springel05,Sijacki07,BoothSchaye09,McCarthy10,Teyssier11,LeBrun14}.

In this work, we study gas and stellar mass fractions in a new set of hydrodynamical simulations of massive haloes.  We find a significant anti-correlation between gas and stellar mass fractions that persists into the deeply non-linear regime.  This anti-correlation does not simply reflect the well-known trends in the mean component fractions with total mass; the anti-correlation exists for {\sl deviations} about the mean trends, meaning it reflects statistical behaviour of the component fractions at {\sl fixed halo mass}.

The new set of simulations is selected from the $N$-body simulation sample {\sc Rhapsody} \citep{Wu13, Wu13b} and re-simulated with gas; we thus name our new sample {\sc Rhapsody-G}.  The original {\sc Rhapsody} sample has been developed with the aim of understanding the impact of formation history on various mass tracers of galaxy clusters.  In this paper, we focus only on the gas and stellar mass of the {\sc Rhapsody-G} clusters; in companion papers (Hahn et al., in preparation and Martizzi et al., in preparation), we will present detailed comparison between our simulations and observational results, including the properties of the BCG and the stellar mass--halo mass relation.

This paper is organized as follows.  
Section~\ref{sec:sim} introduces the simulations.
In Section~\ref{sec:fgas_fstar}, we discuss the anti-correlation between gas and stellar mass fractions, while in Section~\ref{sec:Mbary}, we discuss using total baryon mass as a low-scatter cluster mass proxy.
We discuss the observational implications in Section~\ref{sec:obs} and summarize our results in Section~\ref{sec:summary}.

Throughout this paper, we use radial and mass scales defined by a spherical density contrast with respect to the critical density of the Universe; e.g. $\Rfivec$ indicates the radius within which the average density is 500 times the critical density at the redshift of interest.

\section{Simulations}\label{sec:sim}

The current {\sc Rhapsody-G} simulation suite includes 10 hydrodynamical zoom-in simulations centred on massive haloes from the original {\sc Rhapsody} sample.  Nine are chosen to have similar final mass of $M_{200} \approx 6 \times10^{14}\Msun$ and the tenth has $M_{200} \approx 1.3 \times10^{15}\Msun$.  We use the adaptive mesh refinement code {\sc Ramses} \citep{Teyssier02} and incorporate cooling, star formation, and AGN feedback.  We describe here the simulation methods and recipes, as well as the details of our sample selection.

\subsection{Simulation methodology}\label{sec:method}

First, we briefly summarize the methods used to generate and post-process the simulations.  We kindly refer the reader to Hahn et al.\ (in preparation) for more details.

{\sl Precursor simulations.} 
The galaxy clusters of the {\sc Rhapsody-G} simulations are based on the {\sc Rhapsody} $N$-body simulations \citep{Wu13, Wu13b}, which include 96 cluster-sized haloes of mass $M_{\rm vir} = 10^{14.8 \pm 0.05} \hiMsun$ re-simulated with a mass resolution of $1.3 \times10^8 \hiMsun$.  These haloes have been identified at $z=0$ in a cosmological volume of $1\ h^{-3} {\rm Gpc}^3$ from the {\sc LasDamas} simulation suite.  The 10 {\sc Rhapsody-G} simulations presented here are selected from the full {\sc Rhapsody} sample in such a way that three of the main haloes have extreme concentration, two have an extreme number of subhaloes, and five have approximately the median concentration and typical number of subhaloes.

{\sl Initial conditions.}
We use the {\sc Music} code \citep{HahnAbel11} to generate the initial conditions of the hydrodynamical simulations, at a starting redshift of $50$. The {\sc Music} code implements the second-order Lagrangian perturbation theory to generate displacements and velocities of dark matter particles, and the local Lagrangian approximation to generate a consistent initial density field of baryons on the grid. The Lagrangian volumes for the zoom simulations have been chosen to include a sphere of $8\,h^{-1}{\rm Mpc}$ centred on the clusters at $z=0$, which allows us to study a substantial cosmic volume outside the main halo.

{\sl $N$-body and hydrodynamical methods.}  The {\sc Ramses} code is based on the adaptive mesh refinement technique, which solves the hydrodynamical equations on progressively refined grids.  The hydrodynamical solver is based on a second-order Godunov scheme for ideal gases with an equation of state $\gamma = 5/3.$ High-resolution dark matter particles have a mass of $10^9\Msun$, and the highest spatial resolution is physical 5 kpc (maximal refinement level 18), with a mass-based quasi-Lagrangian refinement strategy.  Due to the added expense of modelling the baryons, this first set of {\sc Rhapsody-G} simulations has eight times lower mass resolution than the original {\sc Rhapsody} $N$-body sample.

{\sl Cooling and star formation.}
Our simulations follow the subgrid cooling model from \cite{SutherlandDopita93}, implemented by \cite{Teyssier11} for {\sc Ramses}.  The star formation rate follows $\dot\rho_* = \epsilon_* (\rho_{\rm gas}/t_{\rm ff})$,  with the star formation efficiency $\epsilon_*$=0.03 and the free-fall time $t_{\rm ff}=\sqrt{3{\rm \pi}/(32G\rho)}$, where $\rho$ is the total mass density.

{\sl AGN feedback.}
We modify the AGN feedback model in \cite{Martizzi12, Martizzi14}, which was based on \cite{BoothSchaye09} and \cite{Teyssier11}.  In this implementation, supermassive black holes are modelled as sink particles, which grow based on mergers and Bondi--Hoyle accretion with a boost factor $\alpha$, with an upper limit set by the Eddington rate.  The thermal energy associated with the accretion is not released until the temperature reaches a certain threshold $T_{\rm min}$.  We choose $\alpha = (n_H/n_*)^2$ when $n_H > n_*$ = 0.1 H/cm$^3$ ($n_H$ is the gas density) and $\alpha=1$ otherwise; $T_{\rm min}=10^7$ K.  The AGN thermal energy is injected into a region of four times the cell size.  We do not implement kinetic AGN feedback associated with jets; it has been shown in \cite{Dubois12} that the kinetic feedback does not significantly affect the bulk gas and stellar mass.

In \cite{Teyssier11} and \cite{Martizzi12, Martizzi14}, the feedback energy was distributed based on a volume-weighted approach, whereas for {\sc Rhapsody-G} we adopt a mass-weighted approach. This implementation results in an effect similar to the quasar mode feedback provided by the radiation pressure of AGN \citep[e.g.][]{Debuhr12}.  In addition, we require that black holes form at the centre of gas clumps with accretion rate $> 30 \rm M_\odot/yr$, and we use the gas clump finding algorithm developed by \cite{Bleuler15}.  As we will show in Hahn et al.\ (in preparation), this implementation results in a halo mass--stellar mass relation in agreement with \cite{Kravtsov14}.

{\sl Post processing.}
We modify the phase-space halo finder {\sc Rockstar} \citep{Behroozi11rs} to include multi-resolution, multi-species particles and gas.  We treat the leaf-cells of the adaptive mesh refinement tree as pseudo-particles, thus allowing for direct integration of all matter components in our haloes.  In this work, we use the dark matter density peak as the centre of the halo, which closely coincides with the stellar mass density peak in most cases.  As we focus on the bulk cluster properties for radius $>R_{2500c}$, the choice of centre does not affect any of the results presented here.

We adopt the same flat $\Lambda$ cold dark matter cosmology as in the  {\sc Rhapsody} simulation.  The cosmological parameters in the simulations are as follows: $\Omega_M$ = 0.25; $\Omega_\Lambda$ = 0.75; $\Omega_b$ = 0.045; $h$ = 0.7. Our cosmic baryon fraction value is $\Omega_b/\Omega_M$ = 0.18, which is slightly higher than the value recently reported by {\em  Planck} \citep[0.155, see][]{Planck13Cosmo}.

\subsection{Sample selection}

\begin{figure}
\includegraphics[width=\columnwidth]{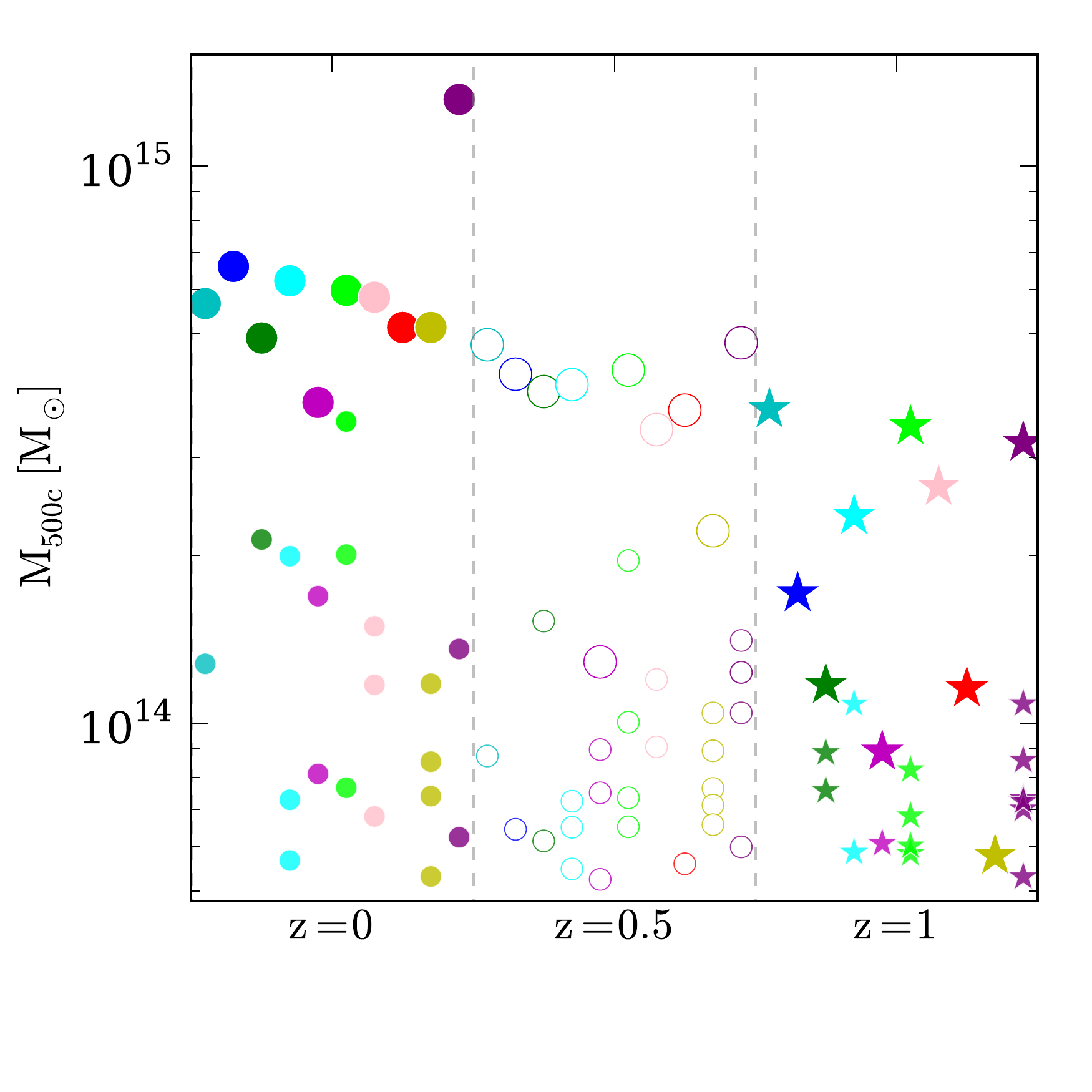}
\vspace{-1.5truecm}
\caption{Mass and redshift of the haloes used in this work.  
Each colour represents one zoom-in simulation centred on a main halo
with $M_{500c}\approx 6\times10^{14}\Msun$ at $z$=0.
We include three snapshots for each simulation: $z$ = 0, 0.5, and 1.
}
\label{fig:sample}
\end{figure}
\begin{table}
\centering
\setlength{\tabcolsep}{0.5em}
\begin{tabular}{ccc|ccc}
\hline
\multicolumn{3}{c|}{Main halo info ($z=0$)} &  \multicolumn{3}{c}{No. of well-resolved haloes}\\ 
\rule[-2mm]{0mm}{6mm} ID &$c_{200}$  &$M_{500c}[\Msun]$& $z=0$ & $z=0.5$ & $z=1$ \\ \hline
\rule[-2mm]{0mm}{6mm} 572 & 7.03 & 5.66$\times10^{14}$ & 2 & 2 & 1 \\
\rule[-2mm]{0mm}{0mm} 337 & 5.19 & 6.61$\times10^{14}$ & 1 & 2 & 1 \\ 
\rule[-2mm]{0mm}{0mm} 377 & 4.79 & 4.91$\times10^{14}$ & 2 & 3 & 3 \\ 
\rule[-2mm]{0mm}{0mm} 348 & 4.62 & 6.22$\times10^{14}$ & 4 & 4 & 3 \\ 
\rule[-2mm]{0mm}{0mm} 653 & 4.47 & 3.77$\times10^{14}$ & 3 & 4 & 2 \\ 
\rule[-2mm]{0mm}{0mm} 361 & 4.41 & 5.98$\times10^{14}$ & 5 & 5 & 5 \\ 
\rule[-2mm]{0mm}{0mm} 448 & 4.40 & 5.81$\times10^{14}$ & 4 & 3 & 1 \\ 
\rule[-2mm]{0mm}{0mm} 545 & 4.40 & 5.13$\times10^{14}$ & 1 & 2 & 1 \\ 
\rule[-2mm]{0mm}{0mm} 211 & 3.65 & 5.13$\times10^{14}$ & 5 & 6 & 1 \\ 
\rule[-2mm]{0mm}{0mm} 474 & 3.55 & 1.32$\times10^{15}$ & 3 & 6 & 7 \\ 
\hline 
\multicolumn{3}{c|}{Totals} & 30 & 37 & 25 \\
\hline
\end{tabular}
\caption{Numbers of distinct haloes in the high-resolution regions with $M_{\rm 500c} >  5\times10^{13}\Msun$.
We sort the list by the concentration of the main halo, $c_{200}=R_{200}/r_s$, at $z=0$.
}
\label{tab:sample}
\end{table}

For this study, we include all haloes in the high-resolution region having $M_{500c}>5\times10^{13}\Msun$ at output redshifts of $z$ = 0, 0.5, and 1. These haloes all satisfy the condition that the mass fraction contributed by low-resolution dark matter particles is below $10^{-3}$.  Since each simulation encompasses a high-resolution sphere of $8\,h^{-1}{\rm Mpc}$ centred on the main halo at $z=0$ and progressively larger high-resolution regions at earlier times, we are able to include 92 haloes in total.

Fig.~\ref{fig:sample} illustrates the masses of the haloes at the three redshifts.  Each colour represents a zoom-in simulation.   For each simulation, the main halo and its most massive progenitors are shown with larger symbols, while other progenitors and nearby high-resolution haloes above the mass threshold are shown with smaller symbols.  The three symbol types correspond to the three redshifts studied here.  The symbol styles will be repeated in the figures below.  Table~\ref{tab:sample} lists the numbers of haloes derived from the outputs of each simulation, with the first column giving the original {\sc Rhapsody} halo ID.

\subsection{Statistical error estimates}

Our main sample consists of all haloes found in the high-resolution regions of the 10 simulations at $z$ = 0, 0.5, and 1.  While the time spacing between these redshifts corresponds to several dynamical times, we do not assume statistical independence among different redshifts for a given halo.   When estimating statistical errors in quantities presented below, we treat the {\sl halo ensemble} extracted from each simulation as independent.  To estimate uncertainties, we jackknife resample using 10 degrees of freedom, eliminating one ensemble of haloes at each round.

\section{Galaxy clusters as nearly closed boxes for baryons}\label{sec:fgas_fstar}
\begin{table}
\centering
\setlength{\tabcolsep}{0.5em}
\begin{tabular}{clc}
\hline
\rule[-2mm]{0mm}{6mm} Symbol & Quantity & Mean ($\Rfivec$)\\ \hline 
\rule[-2mm]{0mm}{0mm} $\fstar$ & Stellar mass fraction & 0.023\\ 
\rule[-2mm]{0mm}{0mm} $\fhot$ & Hot gas mass fraction ($kT> 0.1$ keV) & 0.149 \\ 
\rule[-2mm]{0mm}{0mm} $\fcold$ & Cold gas mass fraction ($kT \le 0.1$ keV) & 0.005\\ 
\hline
\rule[-2mm]{0mm}{0mm} $\fgas$ & Gas mass fraction, $\fgas = \fcold + \fhot$ & 0.154\\ 
\rule[-2mm]{0mm}{0mm} $\fcs$ & Galactic mass fraction, $\fcs = \fstar + \fcold$ & 0.028\\ 
\rule[-2mm]{0mm}{0mm} $\fb$ & Baryon  mass fraction, $\fb = \fstar + \fhot + \fcold$ & 0.177\\ 
\hline
\end{tabular}
\caption{Notation used for baryon mass components.  The first three quantities are derived from the {\sc Ramses} output; the rest are derived from these quantities.  All fractions are relative to the total mass within some chosen density contrast.  
}
\label{tab:notation}
\end{table}
\begin{figure*}
\includegraphics[width=\columnwidth]{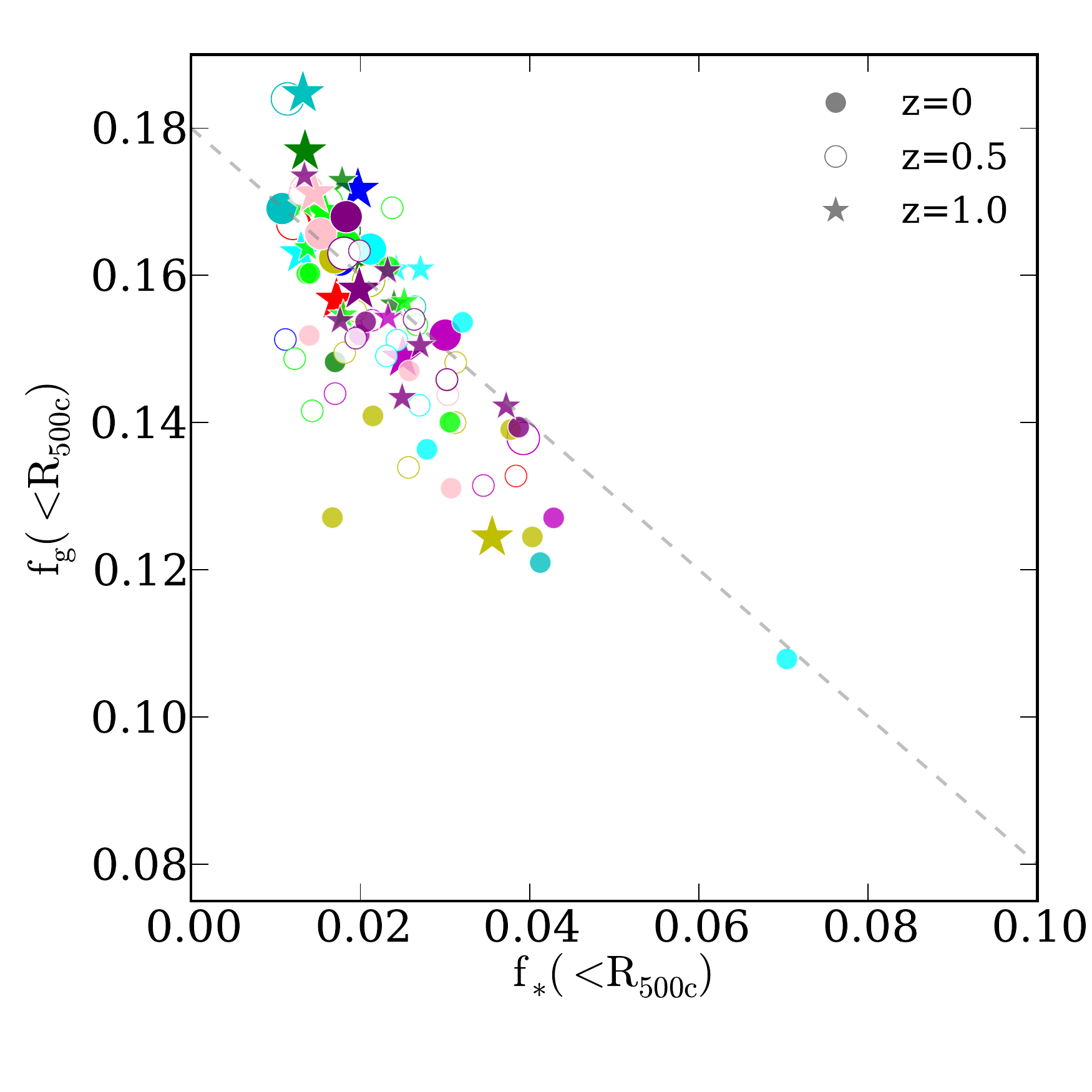}
\includegraphics[width=\columnwidth]{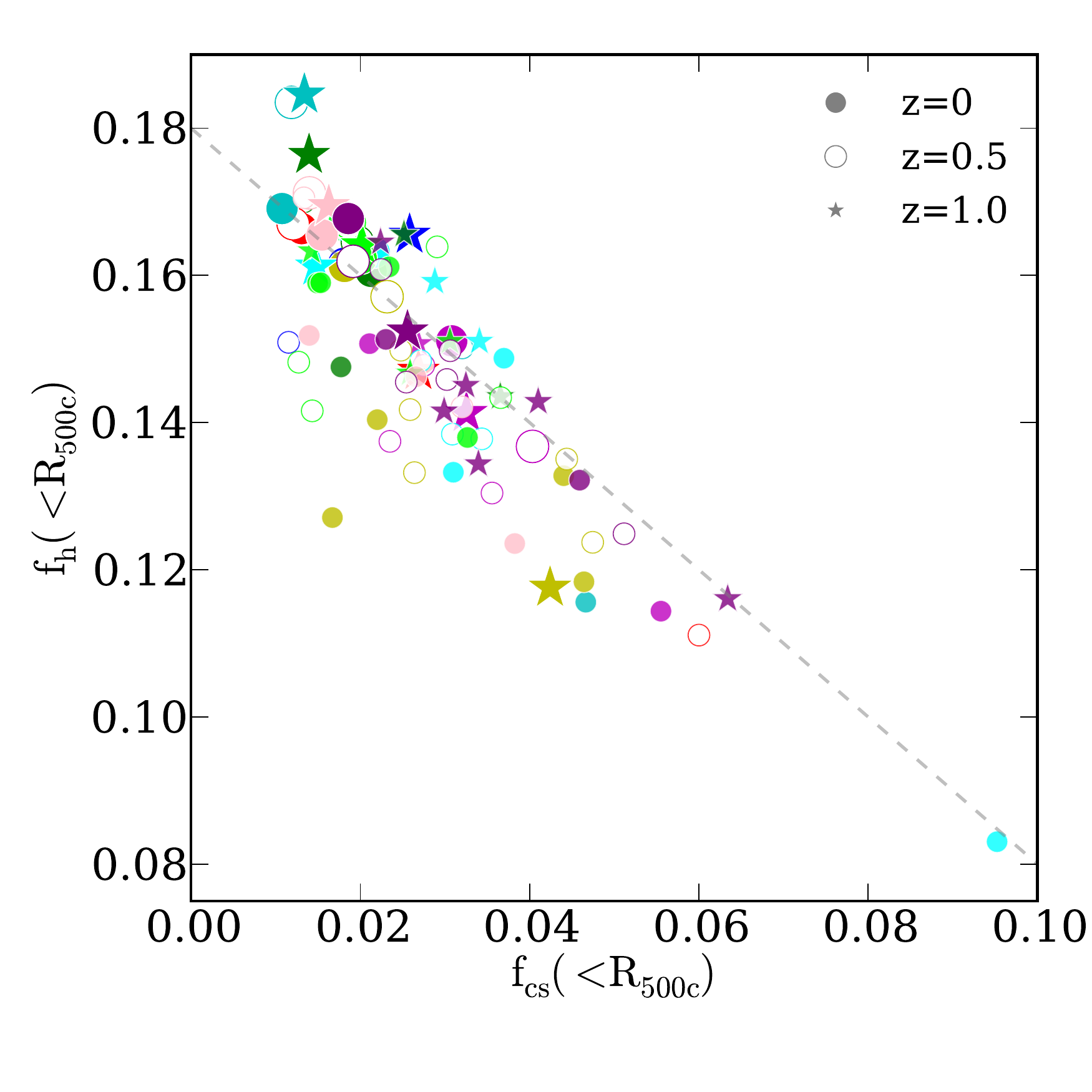}
\vspace{-0.5truecm}
\caption{Anti-correlation between gas and stellar mass fractions.
Left:  the gas mass fraction using both hot and cold phases, $\fgas$, and stellar mass fraction, $\fstar$, 
within $\Rfivec$ exhibit an anti-correlation, $r = -0.72 \pm 0.02$.  Right: the hot  ($kT>$ 0.1 keV) gas fraction, $\fhot$, and galactic (cold gas and stellar) mass fraction, $\fcs$, exhibit a stronger anti-correlation, $r = -0.79 \pm 0.02$.  The symbol styles are the same as used in Fig.~\ref{fig:sample}.  The dashed line in each panel shows the perfect anti-correlation case for which the sum of the two components plotted equals the assumed cosmic mean baryon fraction, $0.18$. 
}
\label{fig:fgas_fstar_cor}
\end{figure*}

In the spherical collapse model of dark matter haloes, the turnaround radius sets the scale within which the cosmic mix of baryonic and cold dark matter should be conserved \citep{GunnGott72}.  Assuming the influence of gravity and collisional shocks, \cite{Bertschinger85} developed self-similar solutions for both collisionless and collisional, ideal fluids, finding that both fluids approached similar radial profiles when expressed in units of the turnaround radius.  Furthermore, the solution for the mixed case of a collisionless fluid with a minority ideal gas component showed no radial separation;  the local interior baryon fraction reflects the cosmic mean value at all radii.

Taking this model to its logical extreme, let us imagine spherical collapse around a local perturbation consisting of radial shells made  of cold dark matter, galactic stars, and smooth intergalactic gas that can shock but is unable to cool.  Collapse of these shells would create a cluster in which the local interior baryon fraction was unbiased at all radii, and in which the mix of stars and gas interior to a given radius would reflect the initial values imposed on the shell layers.

In this section, we show that this naive expectation is respected to a surprising degree for the case in which baryons experience  complex astrophysical processes associated with galaxy formation in a fully three-dimensional, hierarchical clustering environment.

\subsection{Correlations among baryon mass fractions}

For each halo in the sample listed in Table~\ref{tab:sample}, we identify all material within a sphere that encompasses a total mass density contrast $\Delta_c$ with respect to the critical density.  We measure the mass in cold dark matter and in all baryonic components to derive the total mass.  We examine the six baryon mass fractions listed in Table~\ref{tab:notation}.   Hot and cold gas phases are defined using a temperature cut of $kT=0.1$~keV, which is approximately the threshold of X-ray emission.  At $z=0$, the cold gas fraction $\fcold$ is generally very small.

Fig.~\ref{fig:fgas_fstar_cor} shows the behaviour of different baryon mass fractions measured within $\Rfivec$.  The point colouring and styling is the same as used in Fig.~\ref{fig:sample}.   In each panel, the dashed line gives the simple expectation in which the sum of the two components plotted equals the cosmic mean, $\Omega_b/\Omega_m$ (component correlation coefficient of $-1$).

The left panel plots gas mass fraction, $\fgas$, against stellar mass fraction, $\fstar$.  The two components have a rank correlation $-0.72\pm0.02$, where the error bar is calculated by jackknife resampling by removing one of the simulation sets in turn.  In the right panel, we shift the mass in cold gas within $\Rfivec$ to the stellar component, and plot the hot gas fraction, $\fhot$, against the total galactic fraction, $\fcs$.   This split, which more closely represents material inside and outside of galaxies, leads to a stronger anti-correlation of $-0.79\pm0.02$.

These anti-correlations are not entirely surprising, given that the cold gas and stellar mass originated from the cooled hot-phase gas.  Nevertheless, we note that $\fgas$ and $\fstar$ differ by almost an order of magnitude, and that the dynamics within $\Rfivec$ are strongly non-linear and different for collisional and collisionless components. Therefore, such a high degree of correlation is a non-trivial finding.

Some of this anti-correlation is driven by trends in mean baryonic content with halo mass.   Massive clusters are observed to have higher $\fgas$ and
lower $\fstar$ than lower-mass systems, reflecting a lower star formation efficiency in higher mass haloes \citep[e.g.][]{Gonzalez07,Giodini09,Andreon10,Zhang11,Lin12,Gonzalez13,Chiu14}.  As we will show below for our simulations, models with AGN feedback are capable of reproducing these trends \citep[also see, e.g.][]{McCarthy11,Planelles13, LeBrun14}.

We fit the mean dependence of $\fgas$ and $\fstar$ with halo mass and remove these trends to examine correlations between the residuals, $\delta \fgas = \fgas - \avg{\fgas | M}$ and $\delta \fstar = \fstar - \avg{\fstar | M}$.  The mean behaviour is derived from a logarithmic fit.  The correlation coefficients within $\Rfivec$ decline slightly, to $-0.63 \pm 0.02$ and $-0.69 \pm 0.01$ for the left and right panels of Fig.~\ref{fig:fgas_fstar_cor}, respectively.  In Table~\ref{tab:scaling}, we list rank correlation coefficients at various scales with the mass trends removed.

It is interesting to ask whether $\fgas$ is also correlated with other galaxy properties.  We define the stellar mass associated with individual galaxies as the mass within the isophotal contour of 25 mag/arcsec$^2$ in the V band, measured along the direction of the angular momentum. In our simulations, the total stellar mass inside $\Rfivec$ is strongly correlated with the stellar mass of the brightest central galaxy (BCG).  Therefore, unsurprisingly, $\fgas$ and the stellar mass of the BCG are also significantly anti-correlated, with a slightly weaker coefficient of $-0.67\pm0.03$.

On the other hand, the number of galaxies, or specifically the ratio between galaxy number and cluster mass ($N_{\rm gal}/M_{\rm tot}$), has a much weaker anti-correlation with $\fgas$ ($r=-0.42\pm0.04$).  Here $N_{\rm gal}$ is the number of galaxies with stellar mass above $10^{11}\Msun$ within $\Rfivec$; we have tested galaxy stellar mass thresholds between $10^{11}$ and $10^{12}\Msun$, and the correlation stays approximately constant.  The main reason is that the lower-mass haloes in our sample are more strongly dominated by the BCG; that is, although they have  higher stellar mass fraction, their $N_{\rm gal}/M_{\rm tot}$  is still low.   Nevertheless, if we focus on a narrow mass range, $N_{\rm gal}/M_{\rm tot}$ is expected to correlate with $\fstar$ and thus anti-correlate with $\fgas$.  When we consider the most-massive haloes in each snapshot, we find a slightly stronger correlation ($r=-0.49\pm0.07$).

We note that the stellar mass unassociated with individual galaxies, the intracluster light (ICL), is difficult to observe, because it requires observations with high sensitivity at very low surface brightness.  Here we consider the stellar mass associated only with the BCG and satellite galaxies, within the 25 mag/arcsec$^2$ isophote mentioned above.  When we sum all the stellar mass associated with galaxies with stellar mass  $ > 10^{11.5}\Msun$ inside $\Rfivec$, we still find a significant anti-correlation between $\fgas$ and this ICL-excluded $\fstar$  ($-0.69$). This anti-correlation also weakly depends on the galaxy stellar mass threshold.  However, we find that this anti-correlation is largely driven by the BCG; when we exclude the stellar mass of BCG, this anti-correlation is largely diminished ($-0.3$).

We caution that the hydrodynamic and gravitational resolution of $\sim 5$ kpc in our simulations is insufficient to reliably model the ICL component.  Using the 25 mag/arcsec$^2$ threshold, we find an average ICL mass fraction of 55\% of the stellar mass.  This fraction is generally higher than observed, reflecting our models' inability to resolve the half-light radii of all but the brightest galaxies.  A simulation suitable for studying ICL will require higher spatial resolution.  Nevertheless, the consistency of the gas--stellar correlation coefficients derived with and without the ICL indicates that our results are not being driven by the ICL component.

\subsection{Scale dependence}
\begin{figure*}
\includegraphics[width=2\columnwidth]{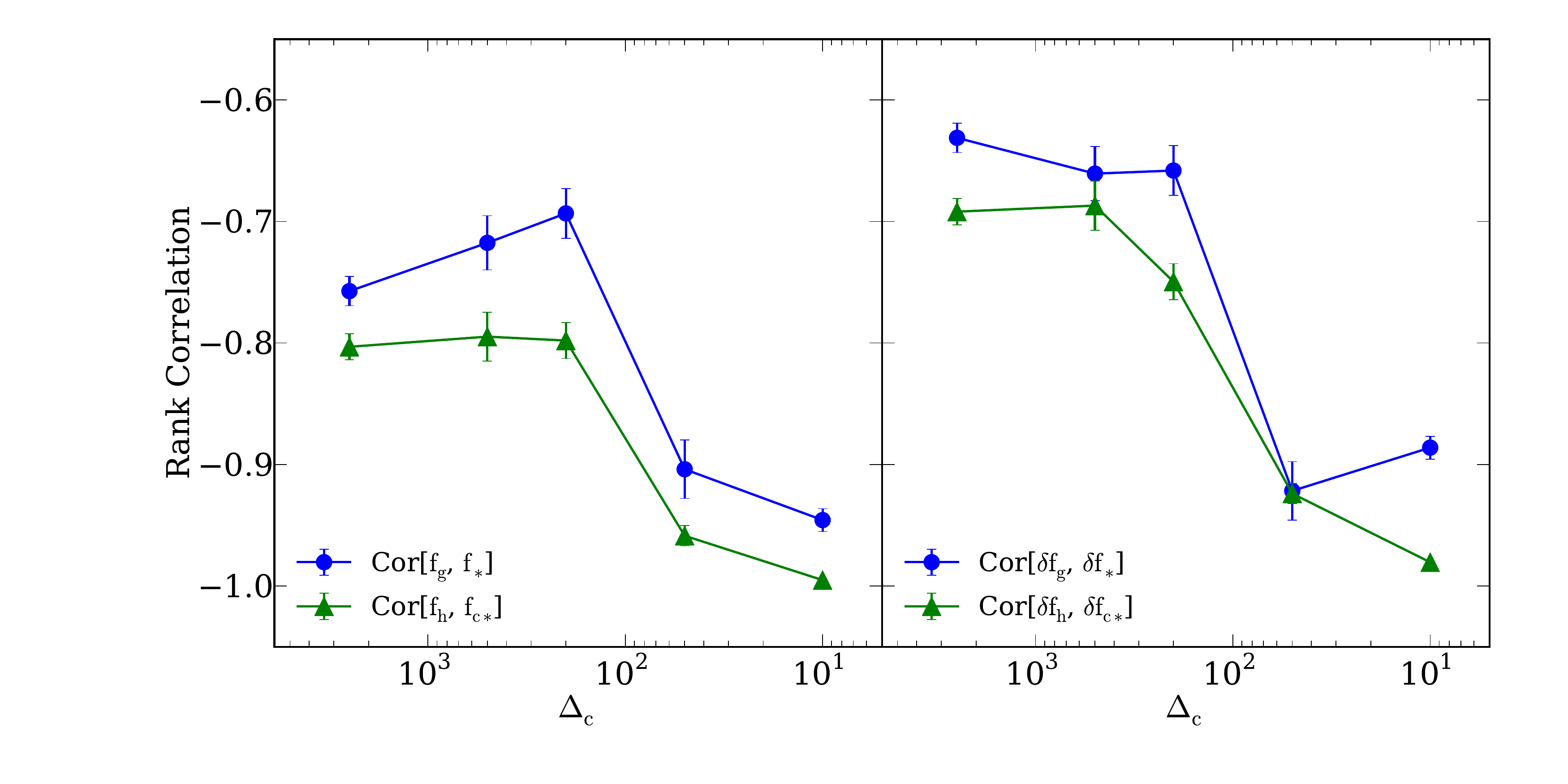}
\vspace{-0.5truecm}
\caption{Anti-correlations between gas and stellar mass fractions as a function of overdensity $\Delta_c$.  Small $\Delta_c$ (large radius) is to the right.  At large radius (low $\Delta_c$), the anti-correlation is significantly stronger, but it remains robust at small radius.   The anti-correlation is stronger between $\delta\fhot$--$\delta\fcs$ (green) than between $\delta\fgas$--$\delta\fstar$ (blue); $\delta f$ is defined as $f-\avg{f|M}$, i.e., with the mass dependence subtracted.}
\label{fig:fgas_fstar_cor_Delta}
\end{figure*}
\begin{figure*}
\hspace{-1cm}
\centerline{\includegraphics[width=2.5\columnwidth]{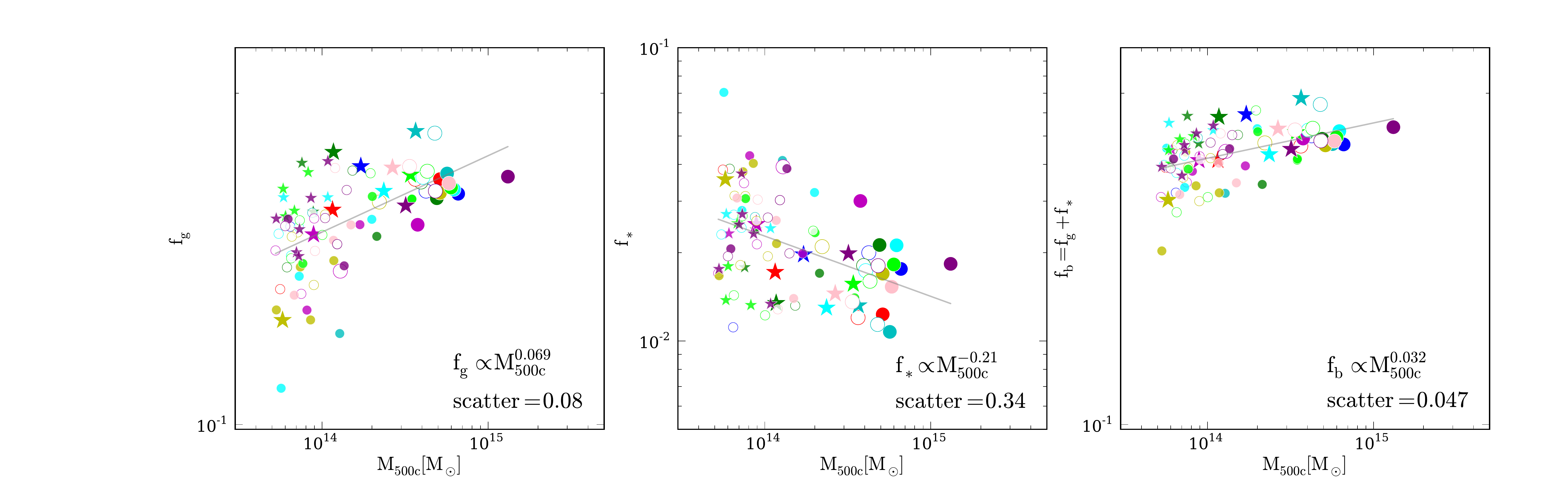}}
\caption{Scaling relations of gas mass fraction (left), stellar mass fraction (middle), and their sum (right) with total halo mass at $z$ = 0, 0.5, and 1.  The symbol styles are the same as used in Fig.~\ref{fig:sample}.  The general trends with mass are consistent with those observed, while the exact values of the slopes may differ from those observed because our sample is incomplete at the low mass end.
}
\label{fig:Mbary}
\end{figure*}

We expand the results shown in Fig.~\ref{fig:fgas_fstar_cor} to explore the scale dependence of baryon component correlations.  We compute statistics within radii that encompass overdensities of $\Delta_c$ = 2500, 500, 200, 50, and 10.  For the two lowest overdensities, we use only the most massive progenitor halo in each simulation to avoid contamination from low-resolution particles.

Fig.~\ref{fig:fgas_fstar_cor_Delta} shows the correlation coefficients as a function of density contrast (large radii are towards the right).  The circular symbols give the correlation between $\fgas$ and $\fstar$, while the triangular symbols correspond to the correlation between $\fhot$ and $\fcs$.  The left panel gives the raw correlation before removing the mass trends, while the right panel removes the effect of the mass trends.  In Table~\ref{tab:scaling}, we list values of the latter.

The anti-correlation between hot phase and galactic (cold gas and stars) components is stronger than that between gas and stars at all radii.  In both panels, the former is very close to $-1$ at $\Delta_c = 10$.  At low density contrasts, the influence of trends with halo mass is weak.   At higher density contrasts, the trends with halo mass are more important, and correcting for them dilutes the raw correlations by $\sim 0.1$. Still, within $R_{2500c}$ the correlation between hot and total galactic baryon residuals  is $-0.71 \pm 0.02$.   This strong covariance indicates that the simple closed-box model remains approximately valid even at radii where complex galaxy formation physics is at play.

\subsection{Dependence on the astrophysical treatment}

We caution that the anti-correlation between $\fgas$ and $\fstar$ presented above is based on one particular astrophysical treatment of star formation and feedback.  To test whether these results are sensitive to this treatment, we employ a set of 51 cluster simulations from \citet[][M14 hereafter]{Martizzi14}.  These simulations are based on the same code and methods as used here, but they employ a volume-weighted AGN energy injection model as opposed to the mass-weighted model in {\sc Rhapsody-G}  (see Section~\ref{sec:sim}).  With the volume-weighted feedback, energy injection within the core is more efficient and the BCG star formation is more strongly suppressed compared to mass-weighted feedback.

The M14 sample reproduces many of the trends presented in this paper, but there are some differences.  In particular, the stellar mass fraction, while similar in the mean to that of {\sc Rhapsody-G}, has a factor of 2 smaller scatter at a fixed halo mass ($17\%$ in M14 compared to $34\%$ in {\sc Rhapsody-G} at $\Rfivec$). 
The observed value reported by \citet{Kravtsov14} of $0.29\pm0.09$ (fractional scatter, not dex) slightly favours the latter but is marginally consistent with both.

Along with the smaller scatter in $\fstar$ at $\Rfivec$, the anti-correlation between $\fgas$ and $\fstar$ at a given mass in the M14 sample is also reduced by a factor of 2, to $r = -0.35$.  The anti-correlation thus persists qualitatively, but the M14 sample behaves somewhat less like `closed boxes' within their non-linear regions than do the {\sc Rhapsody-G} sample.

The small $\fstar$ scatter in the M14 sample is associated with the implementation of the AGN feedback.  The exact value of the correlation coefficient between $\fgas$ and $\fstar$ thus depends on how AGN feedback is modelled.  Uncertainties in predicting stellar masses in simulations \citep[e.g.][]{Ragone13,Martizzi14} imply concurrent uncertainties in the correlation between $\fgas$ and $\fstar$.  Measuring the anti-correlation between hot gas and galactic mass fractions, discussed in Section~\ref{sec:obs}, thus offers a means to constrain details of the feedback model.

\section{Baryon mass as a low-scatter proxy of total mass}\label{sec:Mbary}
\begin{table*}
\centering
\textbf{Summary of correlation and scatter for gas, stellar, and baryon mass fractions}\par\medskip
\setlength{\tabcolsep}{0.5em}
\begin{tabular}{c|cc|ccc}
\hline \hline
\rule[-2mm]{0mm}{6mm}  &\multicolumn{2}{c|}{Rank correlation}  & \multicolumn{3}{c}{Fractional scatter}   \\ 
\rule[-2mm]{0mm}{0mm}  & $(\delta\fgas$, $\delta\fstar$) & ($\delta\fhot$, $\delta\fcs$) & 
$\fgas$ & $\fstar$ & $\fbary$  \\
\hline
\rule[-2mm]{0mm}{5mm} $R_{2500c}$ & $-0.63 \pm 0.02$ & $-0.69 \pm 0.01$ & 0.19$\pm$0.005 & 0.38$\pm$0.007 & 0.11$\pm$0.002 \\  
\rule[-2mm]{0mm}{5mm} $\Rfivec$ & $-0.66 \pm 0.02$ & $-0.69 \pm 0.02$ & 0.08$\pm$0.002 & 0.34$\pm$0.005 & 0.047$\pm$0.002 \\  
\rule[-2mm]{0mm}{5mm} $R_{200c}$ & $-0.66 \pm 0.02$ & $-0.75 \pm 0.02$ & 0.062$\pm$0.002 & 0.32$\pm$0.005 & 0.038$\pm$0.001 \\  
\rule[-2mm]{0mm}{5mm} $R_{50c}$ & $-0.92 \pm 0.02$ & $-0.92 \pm 0.01$ & 0.024$\pm$0.002 & 0.26$\pm$0.02 & 0.01$\pm$0.0007 \\  
\rule[-2mm]{0mm}{5mm} $R_{10c}$ & $-0.89 \pm 0.03$ & $-0.980 \pm 0.003$ & 0.015$\pm$0.001 & 0.22$\pm$0.02 & 0.0047$\pm$0.0003 \\  
\hline
\end{tabular}
\caption{
Correlation coefficients between gas and stellar mass fraction (second column), between hot gas and galactic baryon mass fraction (third column), as well as the fractional scatter in gas mass (fourth column), stellar mass (fifth column), and baryon mass (sixth column) fractions, at radii characterized by different overdensities.}
\label{tab:scaling}
\end{table*}

The  anti-correlations in baryonic components shown above reflect the fact that the sum of these components has lower scatter with respect to total halo mass than does each component.  Fig.~\ref{fig:Mbary} shows scaling relations for gas mass, stellar mass, and baryon mass fractions as a function of total halo mass, measured within $\Rfivec$.  The left panel shows that $\fgas$ is very tightly correlated with the total mass, with a scatter of $8\%$ and a slope of $0.08$.  We note that this scatter is slightly lower than previously reported using a wider range of halo masses; e.g. \cite{Kravtsov06} reported a fractional scatter of $0.107$ (based on $\Rfivec$, see their table 2).  On the other hand, our scatter is similar to the observed value recently reported by \cite{Mantz14}.  For a sample of 40 relaxed clusters observed with {\em Chandra},  they have found $7\%$ scatter in $\fgas$ within a shell of $0.8$--$1.2R_{2500c}$.

The middle panel shows that $\fstar$ has a much larger scatter of 34\% (0.13 dex) and a slope of $-0.21$.  In observations, \cite{Kravtsov14} recently reported that the $\Mstar$--$M_{500c}$ relation has a similar scatter (0.11$\pm$0.03 dex). The right panel shows that when $\fgas$ and $\fstar$ are combined to form $\fbary$, the scatter is 4.7\%, which is smaller than using either $\fgas$ or $\fstar$ alone.  The overall trends with mass reflect that low-mass haloes have higher efficiency of turning gas into stars, but at the same time they tend to have larger baryon depletion.

\begin{figure}
\includegraphics[width=\columnwidth]{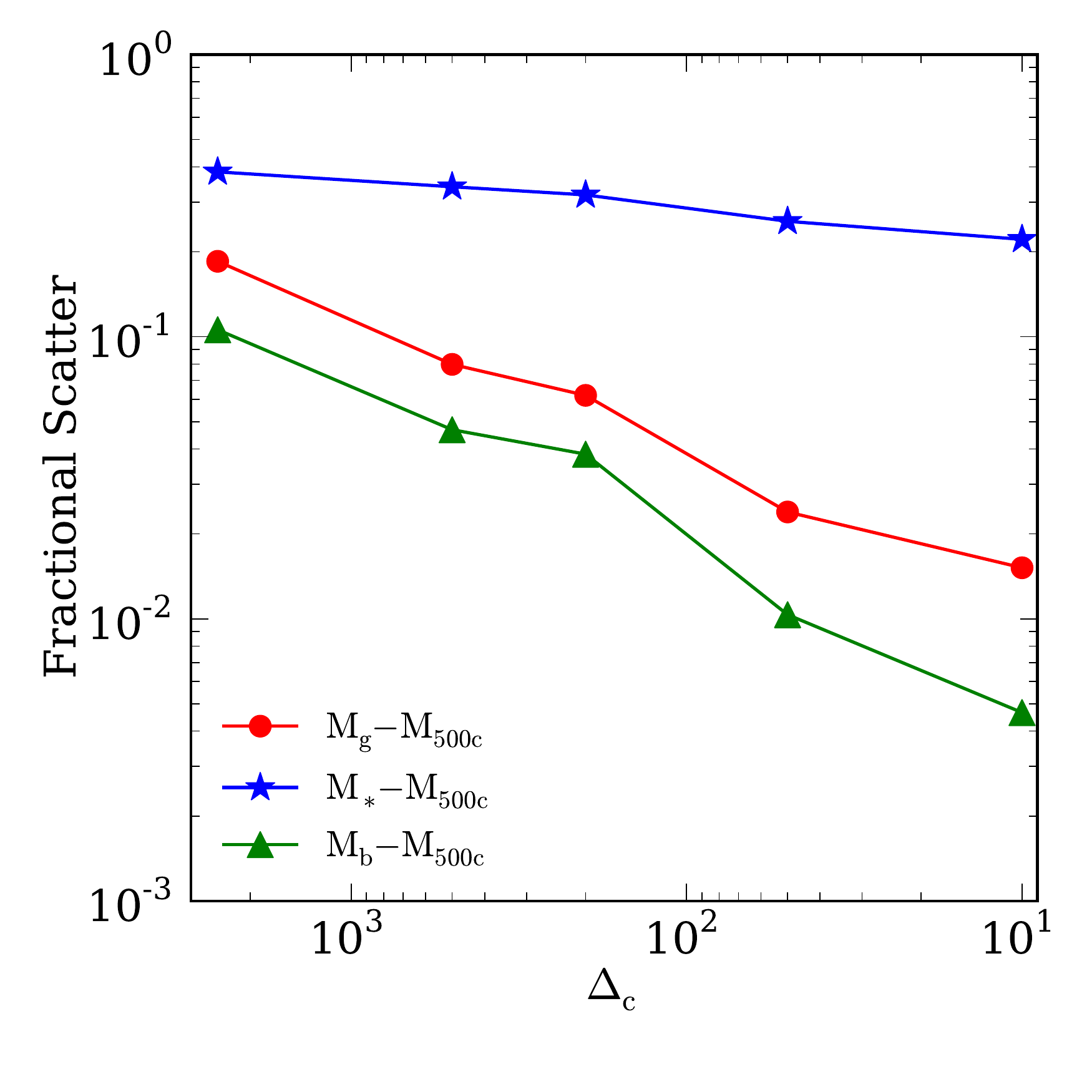}
\vspace{-0.5truecm}
\caption{Scatter in baryon mass proxies as a function of radius defined by the enclosed overdensity, $\Delta_c$.  The $\Mbary$--$\Mtot$ relation has the lowest scatter at all radii, falling to $0.5\%$ at $\Delta_c=10$.}
\label{fig:Mbary_scatter_Delta}
\end{figure}

In Fig.~\ref{fig:Mbary_scatter_Delta} we show the fractional scatter of the different baryon mass components as a function of scale.  The fractional scatter of $\Mstar$ is the largest.  The scatter in $\Mgas$ is smaller and declines more rapidly with increasing radius.  The fractional scatter in the overall baryon mass, $\Mbary$, is typically a factor of 2 smaller than that of the gas mass, declining to $0.5\%$ at  $\Delta_c=10$.  Overall, the baryon fraction scatter scales approximately as $ \langle (\delta \fb/\fb)^2 \rangle^{1/2} \simeq  0.018  (\Delta_c/100)^{0.58}$.

The reduced scatter in the total baryon mass can also be understood in the context of joint property selection.  Under the assumption of a joint lognormal distribution for two observables at a fixed halo mass, the scatter of halo mass obtained by joint selection in these observables is given by \citep{Evrard14}
\beq
\sigma_{\rm joint}^2 = \frac{1-r^2}{\sigma_1^{-2} + \sigma_2^{-2} - 2 r \sigma_1^{-1} \sigma_2^{-1}} , 
\label{eq:joint}
\eeq
where $\sigma_1$ and $\sigma_2$ are the fractional scatter in the two observables at a fixed mass, $r$ is their correlation coefficient, and $\sigma_{\rm joint}$ is the resultant scatter in halo mass under joint selection.   
Evaluating this expression using $\sigma_1 = 0.08$, $\sigma_2 = 0.34$, and $r = -0.72$, we obtain $\sigma_{\rm joint} = 0.047$, which matches the fractional scatter we have found in $\Mbary$.

\section{Observational implications}\label{sec:obs}

In this work, we predict a significant anti-correlation between gas and stellar mass fractions in the virial regions of galaxy clusters.  Here, we discuss a few possible approaches that could be applied to current and future observations to test for this signature.  The fundamental challenge lies in understanding sources of noise 
that would drive the measured covariance away from the intrinsic value.

\subsection{Individual measurements of $\fgas$ and $\fstar$}\label{sec:direct}

A direct approach is to measure gas, stellar, and total masses for an ensemble of clusters and examine the correlations in the  component fractions.  Ideally, the noise in such measurements would be smaller than the intrinsic scatter, which is approximately 10\% for $\Mgas$ and 30\% for $\Mstar$.

Gas masses can be determined fairly reliably;  \cite{Rozo14a} showed that, after accounting for aperture biases driven by total mass errors, gas mass estimates of individual clusters deviate in the mean by just a few percent across three independent observing teams using different X-ray telescopes and analysis methods.  Statistical uncertainties from photon statistics can be made small for estimates at $\Delta_c \ge 500$ in low redshift clusters.

Deriving the stellar mass of clusters from optical images is more difficult.  Stellar masses of bright galaxies are sensitive to the surface brightness profiles fit to the photometry \citep[e.g.][]{Bernardi13}, and the revised fits by \cite{Kravtsov14} pushed BCG stellar masses of SDSS galaxies up by factors of 2--4.  Systematic uncertainties in cluster membership assignment can produce errors in the stellar mass contributed by satellite galaxies \citep[e.g.][]{Lin04}, and uncertainties in the stellar population synthesis models can also contribute  \citep[e.g.][]{Conroy09}.  Finally, measuring and defining intracluster light involve additional uncertainties \citep[e.g.][]{LinMohr04,Gonzalez07,Budzynski14}.

These additional uncertainties would dilute the measured correlation.   Adding random fractional error, $\sigma_0$, to the stellar mass estimates alone would decrease the correlation coefficient by a factor of $\sigma_\ast / \sqrt{\sigma_0^2 + \sigma_\ast^2}$, where $\sigma_\ast$ is the intrinsic scatter of roughly $30\%$ (Table~\ref{tab:scaling}).  For $\sigma_0 = \sigma_\ast$, the correlation coefficient would reduce to $\sim -0.5$.

Statistical error of a few tens of percent is expected in total mass estimates of individual clusters derived from either X-ray hydrostatic equilibrium  \cite[e.g.][]{Rasia06,Nagai07,Rasia12,Rasia14,Nelson14} or weak gravitational lensing \citep[e.g.][]{BeckerKravtsov11,OguriHamana11,Bahe12,Rasia12}.  However, \citet{Donahue14} have provided evidence that combined strong and weak gravitational lensing models reduce the scatter in total mass estimates.  Since error in total mass induces a shift of the same sign for $\fgas$ and $\fstar$, such errors would smear out the trend seen in Fig.~\ref{fig:fgas_fstar_cor} by introducing scatter perpendicular to the anti-correlation.  A potential alternative is to avoid total mass estimates and select instead on a low-scatter mass proxy, such as X-ray temperature or the product between gas mass and X-ray temperature ($Y_{\rm X}$).  One could then examine how $\Mgas$ and $\Mstar$ within fixed metric apertures covary within that sample.  This approach would take advantage of the relatively  weak radial dependence of the correlations exhibited in Fig.~\ref{fig:fgas_fstar_cor_Delta}.

\begin{figure}
\includegraphics[width=\columnwidth]{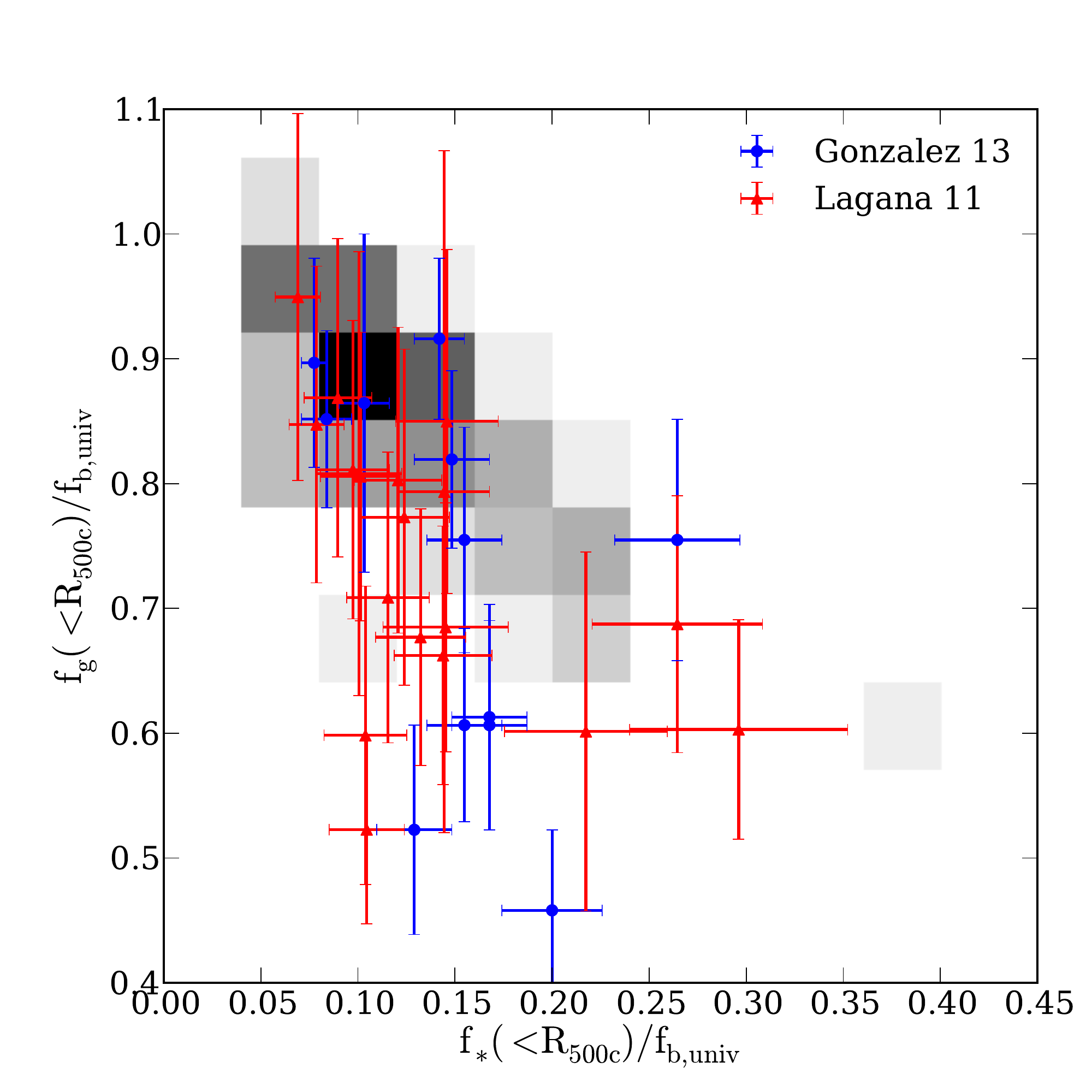}
\caption{Comparison of component mass fractions in our simulations (grey 2D histogram) with observed clusters from \protect\cite{Lagana11} and \protect\cite{Gonzalez13}.  The component fractions are normalized to the universal baryon fraction, $f_{b,univ} \equiv \Omega_b/\Omega_m$, which for our simulations is $0.18$.  For the observations, we use the recent {\em Planck} satellite value of $0.155$.}
\label{fig:fgas_fstar_with_obs}
\end{figure}

Fig.~\ref{fig:fgas_fstar_with_obs} compares our results with several observed clusters in the literature.  Our simulations are represented by the grey histogram in the background (the same data points as in the left panel of Fig.~\ref{fig:fgas_fstar_cor}).  We include observed clusters with $M_{\rm 500c} > 5\times10^{13}\Msun$
published in \cite{Lagana11} and \cite{Gonzalez13}.   We note that these data sets are based on different mass calibration techniques.  \cite{Lagana11} used a $\Mgas$--$M_{500c}$ relation to calculate $M_{500c}$  and used the Schechter function for $\Mstar$; \cite{Gonzalez13} used a $T_X$--$M_{500c}$ relation to calculate $M_{500c}$  and carefully accounted for ICL when calculating $\Mstar$.

In our simulations, $\Omega_b/\Omega_m = 0.18$, which is higher than the value of $0.155$ recently constrained by the {\em Planck} satellite \citep{Planck13Cosmo}.   
To account for this difference, we normalize the simulated component fractions by our $\Omega_b/\Omega_m$, and the observed fractions by the {\em Planck} value.
As shown in Fig.~\ref{fig:fgas_fstar_with_obs}, the statistical error bars on $\fgas$ and $\fstar$ are fairly large, and there are systematic effects that we cannot correct for.  Setting these caveats aside, the simulated and observed trends in Fig.~\ref{fig:fgas_fstar_with_obs} are roughly consistent.

\subsection{Multi-property scaling relations}

Since directly measuring the correlation between $\fgas$ and $\fstar$ is challenging, a practical alternative is to consider statistical effects on scaling relations for property-selected samples.  The basic idea is as follows:  if we select a subset of clusters based on property $S_a$, then the mean and scatter of a second property, $S_b$, will depend on how $S_a$ and $S_b$ are correlated at a given mass.  Following the formalism of \cite{Evrard14}, we assume a power-law mass--observable scaling, so that 
\beq
\big\langle s_a \big| \mu \big\rangle  =  \pi_a + \alpha_a \mu ,
\eeq
where $s_a = \ln S_a$, $\mu = \ln M$, and $\pi_a$ gives the normalization in the chosen units.  Denoting the scatter in $s_a$ at a given mass by $\sigma_a$, then the scatter in $\mu$ at a given $s_a$  is  $\sigma_{\mu|a} = \sigma_a / \alpha_a $.

The mass scaling of the second observable, $S_b$, follows similar notation.  Let $r_{ab}$ be the correlation coefficient of $s_a$ and $s_b$ at a fixed mass.  Finally, the convolution from mass to observed signal requires knowledge of the mass function, which can be approximated as $n(M) = A e^{-\beta \mu}$, with $-\beta$ the local logarithmic slope of the mass function.

For a sample selected on observable $s_a$, the mean value of $s_b$ is given by
\beq
\big\langle s_b \big| s_a \rangle = \pi_b + \alpha_b  \bigg[
\big\langle \mu \big| s_a \big\rangle  + \beta \ r_{ab}\ \sigma_{\mu|a}\ \sigma_{\mu|b}
\bigg]  \ ,
\label{eq:mean}
\eeq
where $\big\langle \mu \big| s_a \big\rangle$ is the mean halo mass selected by $s_a$.  
The variance of $s_b$ is given by
\beq
\sigma_{b|a}^2 = 
\alpha_b^2 \left[  {\sigma_{\mu|a}}^2 + {\sigma_{\mu|b}}^2 - 2 \ r_{ab} \ \sigma_{\mu|a} \ \sigma_{\mu|b} \right]
\label{eq:var}
\eeq
These expressions show that if $r_{ab} < 0$, then the mean of $s_b$ will be biased low, and its scatter will be larger compared to the case of no correlation.

The effect on the mean will generally be small.  For example, if the two observables have 20\% mass scatter, and $\beta \simeq 2.5$, then the final term in Equation~\ref{eq:mean} is a shift of $0.1 r_{ab}$ in $\ln M$.  Such small shifts, below 10 percent in mass, are currently challenging to measure, since the systematic errors in mass are of similar or larger magnitude. In addition, measuring this shift requires accurate knowledge of the observable--mass normalization, $\pi_b$, as well as the mean selected mass.

Analysing the variance is a potentially simpler alternative.  If properties $a$ and $b$ have comparable mass scatter, then all the terms on the right-hand side of Equation~\ref{eq:var} will be of the same order.  As an example, we consider recent observational results involving galaxy richness (as $S_a$) and gas mass ($S_b$) in \cite{RozoRykoff14}, which are summarized in their table 2.  The mass scatter at a fixed richness is approximately $\sigma_{\mu|a} = 0.25$ \citep{Rykoff12}, while the mass scatter at a fixed $\Mgas$ is approximately $\sigma_{\mu|b} = 0.1$ \citep{Mantz10b,Mantz14}.  Based on matching existing X-ray data to the optically selected redMaPPer cluster sample, \cite{RozoRykoff14} report a scatter in gas mass at a fixed galaxy richness of $\sigma_{b|a} = 0.212\pm0.032$, and report a slope for the $\Mgas$--$\Mtot$ relation of $\alpha_b = 0.72 \pm 0.12$.

Evaluating Equation~\ref{eq:var} with these values gives $r_{ab} = -0.28$, a slight hint of an anti-correlation.  However, the exact value of $r_{ab}$ sensitively depends on the slope $\alpha_b$, which is poorly constrained.  Decreasing the slope by its one sigma uncertainty, to $\alpha_b = 0.65$, leads to an estimate of $r_{ab} = -0.68$.  
Similar exercises based on larger samples of homogeneously determined mass estimates will offer a more robust means to test for non-zero covariance in stellar and hot gas mass fractions.

A related test is to probe the variance in total halo mass under joint property selection (see Equation~\ref{eq:joint}).  Using optical and X-ray samples for which both $\Mstar$ and $\Mgas$ are accurately measured, one could first use lensing total masses to estimate the scatter in $\Mtot$ for each property selection.   Fitting the lensing masses in the joint selection of both properties would provide a fundamental plane with variance reduced by an amount given by Equation~\ref{eq:joint}.  Selection effects would need to be carefully modelled in such a study.

\section{Summary}\label{sec:summary}

We present an analysis of the various baryonic mass components -- stars, hot and cold gas -- in a sample of $\sim 100$ massive haloes derived from the {\sc Rhapsody-G} cosmological hydrodynamic simulations. These simulations include state-of-the-art models for gas cooling and star formation, as well as for energy injection through supernovae and AGN. Our findings can be summarized as follows:

\begin{itemize}

\item At a fixed total halo mass, stellar and gas mass fractions are significantly anti-correlated in the non-linear regions of haloes, with $r = -0.66 \pm 0.02$ at $\Rfivec$. 
This correlation is further enhanced if we split gas into hot and cold gas, and correlate the hot gas fraction with the cold gas plus stellar mass fraction ($r=-0.69$ at $\Rfivec$).  

\item Due to this anti-correlation, total baryon mass has a scatter with respect to the total halo mass that is lower than either gas mass or stellar mass.   At $\Rfivec$, the baryon mass has approximately 5\% scatter, suggesting that joint cluster selection using accurate gas and stellar mass measurements can achieve up to 5\% selection in total mass.

\item With increasing radius, the anti-correlation between $\fgas$ and $\fstar$ approaches -1, the closed box expectation, and the baryon mass scatter declines to 0.5\% at $\Delta_c = 10$.  

\end{itemize}

It is currently challenging to accurately measure the anti-correlation between $\fgas$ and $\fstar$ in observations.  Scaling laws with well measured slopes, intercepts, and standard deviations are required for large samples.  To obtain empirical constraints on this correlation in massive clusters, joint survey studies that combine gas mass and stellar mass selection with lensing masses and/or additional independent mass proxies (from X-ray temperatures, $Y_X$, caustic masses, galaxy velocity dispersions, etc.) are needed.

Comparison to {\sc Ramses} simulations that use a different AGN feedback scheme indicates that the results are qualitatively robust but quantitatively dependent on the feedback method.  In light of the different simulation results from different implementations of AGN feedback \citep[e.g.][]{Ragone13,Martizzi14} and between adaptive mesh refinement and smoothed-particle hydrodynamics methods \citep[e.g.][]{Frenk99,Rasia14,Sembolini15}, it is important to address the anti-correlation found here using different simulation techniques, more detailed physical models, and different subgrid models for feedback processes. The closed-box result must hold at sufficiently large radii, but the detailed scale dependence of the covariance in baryon components is likely to exhibit model-dependent features.

\section*{Acknowledgements}
We thank Steve Allen, Adam Mantz, Truong Nhut, Elena Rasia, and Yuanyuan Zhang for useful discussions.  We thank Peter Behroozi for kindly making his {\sc Rockstar-Galaxies} code available to us for further modifications. HW acknowledges the support by the U.S.\ Department of Energy under contract number DE-FG02-95ER40899. OH acknowledges support from the Swiss National Science Foundation through the Ambizione fellowship.  DM acknowledges support from the Swiss National Science Foundation.  RW received support from the U.S.\ Department of Energy contract to SLAC no.\ DE-AC02-76SF0051.  This work was supported by a grant from the Swiss National Supercomputing Centre (CSCS) under project ID s416.
\bibliographystyle{mn2e}
\bibliography{/Users/hao-yiwu/Dropbox/master_refs}
\end{document}